# Superhard Phases of Simple Substances and Binary Compounds of the B−C−N−O System: from Diamond to the Latest Results (a Review)

O. O. Kurakevych

*Bakul Institute for Superhard Materials,
National Academy of Sciences of Ukraine,
vul. Avtozavods'ka 2, Kiev, 04074 Ukraine*

**Abstract**—The basic known and hypothetic one- and two-element phases of the B−C−N−O system (both superhard phases having diamond and boron structures and precursors to synthesize them) are described. The attention has been given to the structure, basic mechanical properties, and methods to identify and characterize the materials. For some phases that have been recently described in the literature the synthesis conditions at high pressures and temperatures are indicated.

## 1. INTRODUCTION

Superhard materials are widely used in the industry for production of various tools, and coatings [1–5]. Diamond, the hardest of superhard materials, is used in modern science and technology thanks to its unique properties [6]. In addition to its high hardness (the Vickers hardness for some faces of a single crystal is up to 170 GPa [7]), high heat conductivity [8], wide bandgap [9], and high mobility of electrons and holes [10] are characteristic of diamond. At the same time diamond easily oxidizes at relatively low temperatures [11] and reacts with iron-group metals [12]. The growing demand for diamond-like superhard single-phase materials for high-temperature electronics [13], electrochemistry [14], and for machining hard alloys and ceramics [15] has stimulated a search for novel superhard phases having a higher thermal and chemical stability than diamond has.

The high-pressure synthesis of artificial diamond in Sweden (1953) [16] and in USA (1954) [17], which became possible owing to the development of new apparatuses and procedures, was a real breakthrough in synthesis of artificial superhard materials, showed the possibility to use high-pressure technique in industry, and stimulated interest in this problem. Four years after the first recognized synthesis of artificial diamond, cubic boron nitride (cBN) was produced at high pressures and was accepted as the second-to-diamond superhard phase [18]. Up to now the high-pressure technique is widely used and is the basic one for a synthesis of new superhard phases and composites, whose hardness is comparable to diamond.

Within the past 10 years a number of novel superhard phases have been synthesized thanks mainly to the advancement of high-pressure technique, which makes it possible to attain a pressure up to 25 GPa on samples of macroscopic sizes. Of these novel materials, the compounds of the B−C−N−O quaternary system occupy a prominent place. All known syntheses of superhard materials having the structure of diamond were realized owing to the development of methods of producing graphite-like phases of the B−C−N, B−C, and C−N systems. Badzian et al. [19] were the first to synthesize boron carbonitride with a graphite-like structure using the method of the thermochemical gas phase deposition (CVD). Later the CVD synthesis of phases with a graphite-like structure like boron carbonitride [20], phases of the C−N system [21], and borated graphites [22, 23] became the basic method to obtain graphite-like precursors of the B−C−N system. It was also proposed to use this method to synthesize some phases of the B−C−N system with the diamond structure. However, diamond-like phases produced by this procedure contain a great number of structural defects and exhibit low processing characteristics as compared to phases produced at high pressures. The use of other procedures, like solid–phase chemical interaction at high and low pressures, also allows one to obtain some phases of the B−C−N and C−N systems with the graphite structure.

After synthesis of superhard cubic boron carbonitrides $cBC_xN$ [26–28], which have the hardness comparable to that of diamond and high processing characteristics, quite recently a new superhard individual phase, cubic $cBC_5$, has been produced [29]. These syntheses have shown that the use of high-pressure technique and graphite-like precursors for the above purposes has considerable promise and that the methods of in situ control are highly efficient in high-pressure synthesis. Superhard composites with extreme hardness, which noticeably exceeds that of polycrystalline diamond, have been produced from graphite-like precursors in addition to individual phases [30–31]. In the literature there are descriptions of the synthesis of another diamond-like $dB_2O$ [32] and cubic $C_3N_4$ [33] phases produced from a graphite-like precursor. However, the results of both studies have not been verified and cast some doubt.

The list of superhard phases of the B–C–N–O system is not exhausted by the phases with a diamond-like structure. Boron is the second hardest element [34], which has the most surprising physicochemical properties among other simple substances [35]. Boron-based compounds form a large group of hard and refractory phases, e.g., $B_6O$, $B_4C$, and recently synthesized $B_{13}N_2$ boron subnitride [36], which have unusual crystalline structures, physical, chemical and mechanical properties thanks to strong covalent bonds of the electron-deficient nature [37, 38]. The combination of such characteristics, like the low density, high hardness, strength, chemical and thermal stability, allows a suggestion that the use of boron and its compounds in engineering holds much promise [5, 39, 40].

To characterize known and novel superhard phases and precursors of them, a number of methods are used, the complex of which makes it possible to define the structural type, composition, and the distribution of atoms within unit cells. As the experimental data on new phases are often interpreted by analogy with already known ones, in the present review in parallel with the phases and their structure and mechanical properties basic characteristic spectra (mainly powder X-ray diffraction (XRD)[1], electron energy loss spectroscopy (EELS)[2], and Raman[3] scattering), which are widely used to verify the structure and composition of respective phases, are given.

The B–C–N–O quaternary system includes basic known covalent diamond-like and boron-like compounds, whose structures are three-dimensional lattices formed by short and strong covalent bonds, which are responsible for a high hardness. A variety of the as yet not synthesized hard and superhard compounds have been predicted for this system [41–45], which makes it an important object of scientific enquiry, especially taking into account the recent results on synthesis of superhard $cBC_5$ [29], cBN/wBN nanocomposite [31] and $\gamma$-$B_{28}$ high-pressure phase [46]. Basic known and hypothetical one- and two-element phases of the B–C–N–O system (both superhard phases having structures of diamond and boron and precursors to synthesize them at high pressures and temperatures) are described in this review.

## 2. CARBON PHASES

Atoms of carbon can form bonds of three types that correspond to valent orbital hybridizations $sp^3$, $sp^2$, and sp. It has been generally accepted (with some exceptions) that each valence state characterizes a certain allotropic modification of carbon. Atoms having $sp^3$ hybridization form three-dimensional lattice of diamond, in which all atoms have tetrahedral coordination. Many polytypes of diamond are known: cubic (3C or diamond proper, Fig. 1a), hexagonal (2H or lonsdaleite), and 8H, 12H, 15R, 16H, 20H, and 21R. Physical and chemical properties of all polytypes are similar. Diamond (3C) crystallizes in the $Fd$-$3m$ space group (cubic syngony, $a = 3.5669$ Å). An X-ray diffraction pattern of diamond exhibits characteristic reflections 111, 220, 311, and 400 (see Fig. 1c). Single mode $\Gamma'_{25}$ (1333 cm$^{-1}$) is active in the Raman spectrum, while doping diamond with other elements brings about the structure disordering and appearance of other active modes. An EELS spectrum is characteristic of carbon in a tetrahedral coordination and is widely used to recognize the $sp^3$ hybridization state of a carbon atom in other phases (see Fig. 1d) [47, 48]. The basic mechanical properties of diamond are listed in Table 1.

In addition to mono- and polycrystalline diamond, the existence of sintered samples of both nanoparticles and nanorods produced by a direct high pressure–high temperature synthesis is described in the literature [49–51]. Whereas the XRD patterns are typical spectra of a diamond-like structure with broadened reflections, Raman spectra strongly differ from those of samples having large grains [50, 51]. According to [52], diamond-like clusters as well as many other nanoparticles have the structure, which cannot be correctly described in the framework of three-dimensional space groups of the classical crystallography. In spite of the variety of theoretical studies in the field [52, 53], the description of the nanoparticle structure still remains an open question, which essentially complicates the interpretation of vibrational spectra.

---

[1] XRD characterizes the crystalline structure and usually allows one to establish the position of atoms in the lattice, though in the case of phases of variable composition the type of atoms in one or other crystallographic position, it is impossible to define because of the proximity of scattering factors of the B–C–N–O system elements.
[2] EELS allows one to determine the phase composition and hybridization of atoms (both in monophase and multiphase systems).
[3] Raman scattering characterizes the space and point symmetry groups that correspond to the phase structure, gives an idea of the lattice dynamics and bonding force, and often provides a possibility to define the structural type.

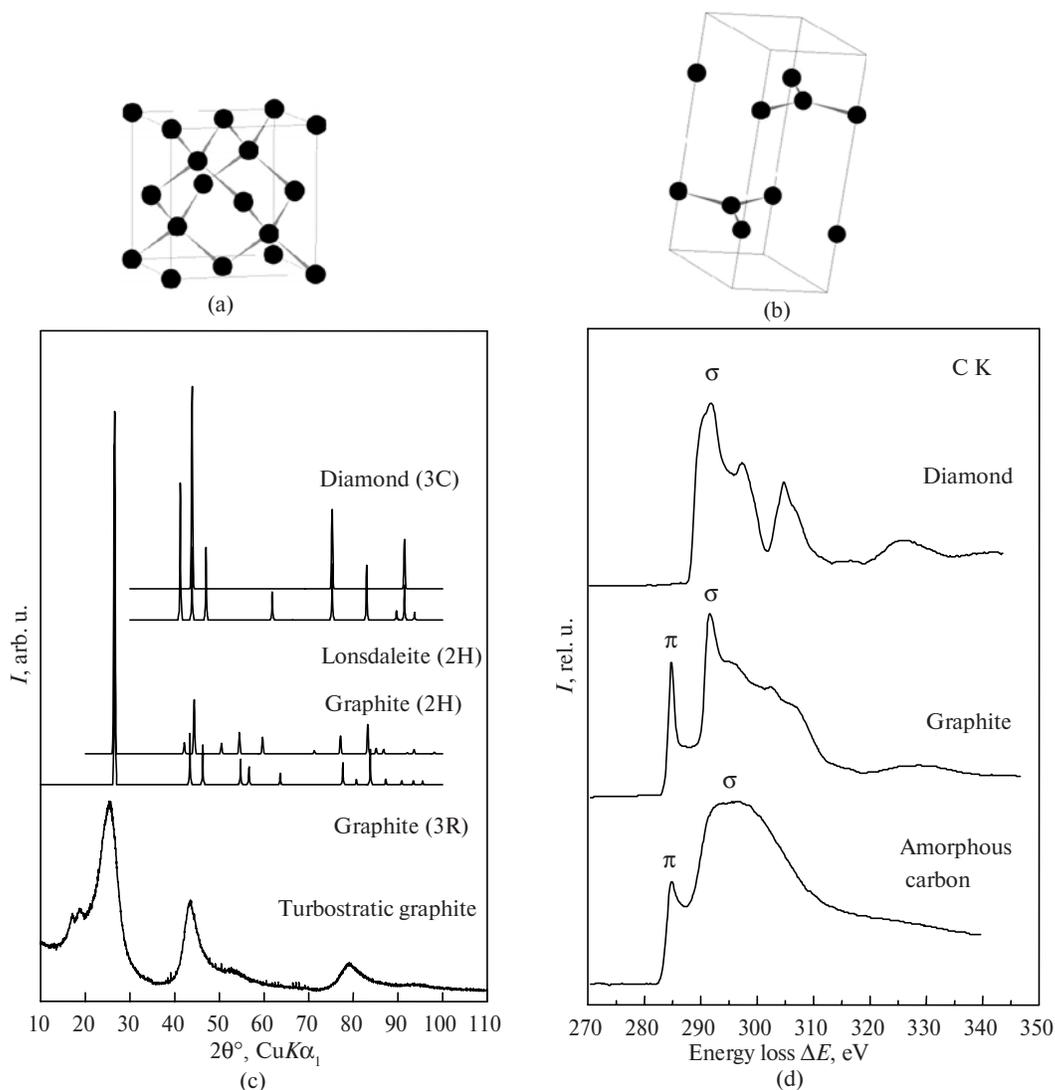

**Fig. 1.** Unit cells of the 3C diamond (a) and 2H graphite (b); X-ray diffraction patterns of diamond, lonsdaleite, graphite (c); EELS spectra of carbon atoms having $sp^3$ and $sp^2$ hybridizations (d).

While graphite is an important precursor for the synthesis of diamond, other graphite-like phases are promising starting materials for the synthesis of new superhard diamond-like phases. Carbon atoms having the $sp^2$ hybridization generate a two-dimensional lattice of graphite layers that form three-dimensional structures of graphite polytypes, like hexagonal graphite (2H or graphite proper), rhombohedral graphite (3R) (see Fig. 1b), and other polytypes (6R, 10H, and 12H). The existence of polytype 1H has been also mentioned in the literature but it has not been verified. The structures of graphites are strongly anisotropic because of the weak Van der Waals forces that bond layers in the direction of the $c$ crystallographic axis as compared to strong bonds within a graphite layer. Hexagonal graphite crystallizes in the $P6_3mmc$ space group (hexagonal syngony, $a = 2.4704$ Å, $c = 6.7244$ Å). XRD patterns of graphite-like structures exhibit characteristic reflex 002 (along with less intensive 004), which corresponds to the interlayer reflection (see Fig. 1c), while the $hkl$ lines correspond to layer reflections. The $hk0$ lines correspond to the mutual ordering of graphite-like layers and the presence of them makes it possible to identify the graphite-like polytypes. In XRD patterns of turbostratic graphite-like phases (the structure is completely disordered in the $c$-axis direction) the $hkl$ reflections are absent and only virtually symmetrical 001 and 002 lines are observed as well as two-dimensional asymmetrical reflexes 10 and 11, which are typical of turbostratic lattices (see Fig. 1c) [54]. Two modes of irreducible representation $E_{2g}$ are active in a Raman spectrum of graphite with an ideal structure (42 and 1581 cm$^{-1}$). The low-frequency mode corresponds to interlayer vibrations, while the high-frequency one comes from intralayer vibrations. The structural disordering often brings about other modes in a spectrum, two lines are observed only in the case of highly ordered natural or pyrolytic samples. A characteristic EELS spectrum allows one to recognize the sp$^2$ hybridization state of carbon atoms in graphite-like phases (see Fig. 1d) [47, 48]. Basic properties of graphite are shown in Table 1.

Monoatomic graphite layers may form nanotubes and spheres (known as fullerenes) [55]. Similar materials are well understood [56–58] and are considered as very promising precursors for synthesis of both superhard sintered diamond nanomaterials [51] and hard polymerized fullerenes [59, 60].

**Table 1.** Properties of basic carbon phases

| Properties | Diamond | Graphite |
|---|---|---|
| Hardness $H_V$, GPa | 115 [61]; 80–130 [62] | |
| Hardness $H_K$, GPa | 90 [63]; 63 [64] | |
| Fracture toughness $K_{Ic}$, MPa·m$^{1/2}$ | 5 [61] | |
| Bulk modulus $B_0$, GPa | 446 [65] | 23 [66] |
| Pressure derivative of $B_0$, $B_0'$ | 4 [65] | 8 [66] |
| Young modulus, $E$, GPa | 1141 [67] | |
| Density $\rho$, g·cm$^{-3}$ | 3.516 | 2.245 |

## 3. MODIFICATIONS OF BORON

The studies of physicochemical properties of boron were started from the moment of its first preparation in 1808 and in the 1950s–1960s boron was rather well understood [68–73]. However, many questions remain unsolved up to now. First of all a very complicated problem is the production of pure boron because impurity enter into its lattice very easily, which often results in heavy structural changes. This is only one of the reasons why the study of structural chemistry of boron required serious effort of many scientists over many decades, which ended with establishing the structure of rhombohedral α-boron by Decker and Kasper [71]. Despite the fact that structures of low-boron borides are relatively simple, they become very complicated with increasing boron content [74], especially in the case of a pure element. At present four allotropic modifications of boron are known: rhombohedral α-B$_{12}$ [71], rhombohedral β-B$_{106}$ [72], tetragonal T-B$_{192}$ [73], and orthorhombic γ-B$_{28}$ [46] (Figs. 2a–2d). A phase diagram of boron to 100 GPa has recently been proposed by Oganov et al. [46] (see Fig. 2e).

The nature of chemical bond and the relations between the structure and properties of boron polymorphous modifications have been detailed in [35, 46]. Usually the structural skeleton of boron consists of icosahedra B$_{12}$, bonded with covalent bonds. The theory of molecular orbitals [35] explains the stability of icosahedra and gives a key to the understanding of the structural chemistry of boron. An interaction between orbitals of boron atoms inside an icosahedron gives rise to 13 bonding, 12 nonbonding, and 23 antibonding orbitals. The electronic configuration of each atom is close to $s^1p^2$ (as opposed to the $s^2p^1$ configuration of a free boron atom) for calculated molecular orbitals [35]. Bonds in boron are of a strongly covalent nature both within an icosahedron and between icosahedra, which is the prerequisite to its high hardness.

Rhombohedral α-B$_{12}$ [71] crystallizes in the *R*-3*m* space group (trigonal syngony, $a$ = 4.927 Å, $c$ = 12.564 Å) (see Fig. 2a). The structure of α-B$_{12}$ is a distorted cubic packing of icosahedra (rhombohedral angle α = 58.06° instead of 60°, as with the ideal packing). The XRD pattern for α-B$_{12}$ is given in Fig. 3a. The active modes in the Raman spectrum correspond to irreducible representation $6E_g + 4A_{1g}$ of the center of the Brillouin zone [75] (see Fig. 3b). The characteristic EELS spectrum for α-B$_{12}$ and phases of this structural type is shown in Fig. 3c [76] and the basic properties of α-B$_{12}$ in Table 2.

Rhombohedral β-B$_{106}$ [72, 77, 78] crystallizes in the *R*-3*m* space group (trigonal syngony, $a$ = 10.9466 Å, $c$ = 23.9034 Å) (see Fig. 2b). The principal base unit of β-B$_{106}$ is polyhedron B$_{84}$, in which 12 pentagonal pyramids B$_6$ surround icosahedron B$_{12}$. The packing of polyhedra, as with α-B$_{12}$, is a distorted cubic packing (rhombohedral angle α = 65.3°). The XRD pattern is shown in Fig. 3a. The Raman spectrum of this modification is rather complicated [79] (see Fig. 3b) because of many active modes: $52E_g$ and $31A_{1g}$. The numerous bands overlap each other, as a result of which the spectrum exhibits broad lines. Also, the situation becomes more complicated due to the presence of partially filled crystallographic positions in the structure. The basic properties of β-$B_{106}$ are given in Table 2.

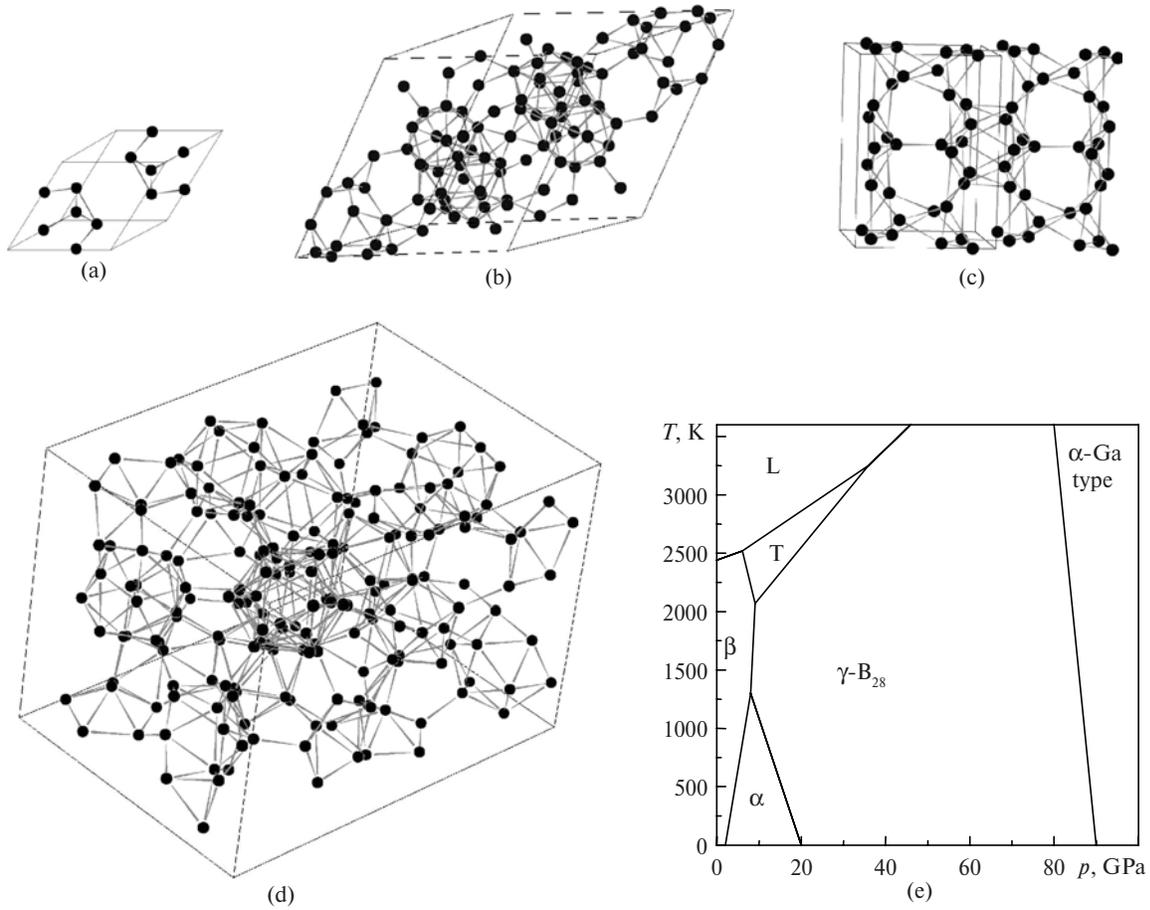

**Fig. 2.** Crystalline structures of boron allotropic modifications: α-$B_{12}$ rhombohedral (a), β-$B_{106}$ rhombohedral (b), γ-$B_{28}$ orthorhombic (c), T-$B_{192}$ tetragonal (d); schematic phase diagram of boron (e).

The studies of α-$B_{12}$ at high temperatures using transmission electron microscopy [80] have shown that in the course of the α-$B_{12}$ to β-$B_{106}$ transformation it is possible to observe and quench two more phases, namely, β′-$B_x$ (1645 K) and β″-$B_y$ (1865 K). According to the structured modeling using the data on electron diffraction of a single crystal, β′-$B_x$ has an eightfold rhombohedral unit cell of the α-$B_{12}$ ($x \approx 96$) type with a rhombohedral angle that exceeds 60° ($a = 10.144$; α = 65.45°), which gives ground to consider this phase as an intermediate one in transition from α-$B_{12}$ (α = 58.06°) to β-$B_{106}$ (α = 65.3°). The β″-$B_y$ structure is hypothetically produced in the course of a systematical shift of boron atoms in β′-$B_x$ (where they followed the α-$B_{12}$ pattern) to form polyhedra typical for β-$B_{106}$ but in the amount of atoms typical of α-$B_{12}$ ($y \approx 96$).

Orthorhombic γ-$B_{28}$ (see Fig. 2c) [46] crystallizes in the *Pnnm* space group (orthorhombic syngony, $a = 5.054$ Å, $b = 5.612$ Å, $c = 6.966$ Å). The phase structure is formed by icosahedra $B_{12}$ and pairs $B_2$, which are located at points of the distorted lattice of the NaCl type. According to the ab initio calculations, a considerable portion of the charge is transferred between $B_{12}$ and $B_2$, as a result of which some B–B bonds in the structure are of partially ionic nature. The XRD pattern is given in Fig. 3 and basic properties are shown in Table 2. This phase has a wide $p$, $T$ region of the thermodynamic stability (see Fig. 2e) and may be produced by quenching at pressures above 10 GPa and temperatures 1800–2000 K. The γ-$B_{28}$ phase is comparable to cubic boron nitride in hardness [34].

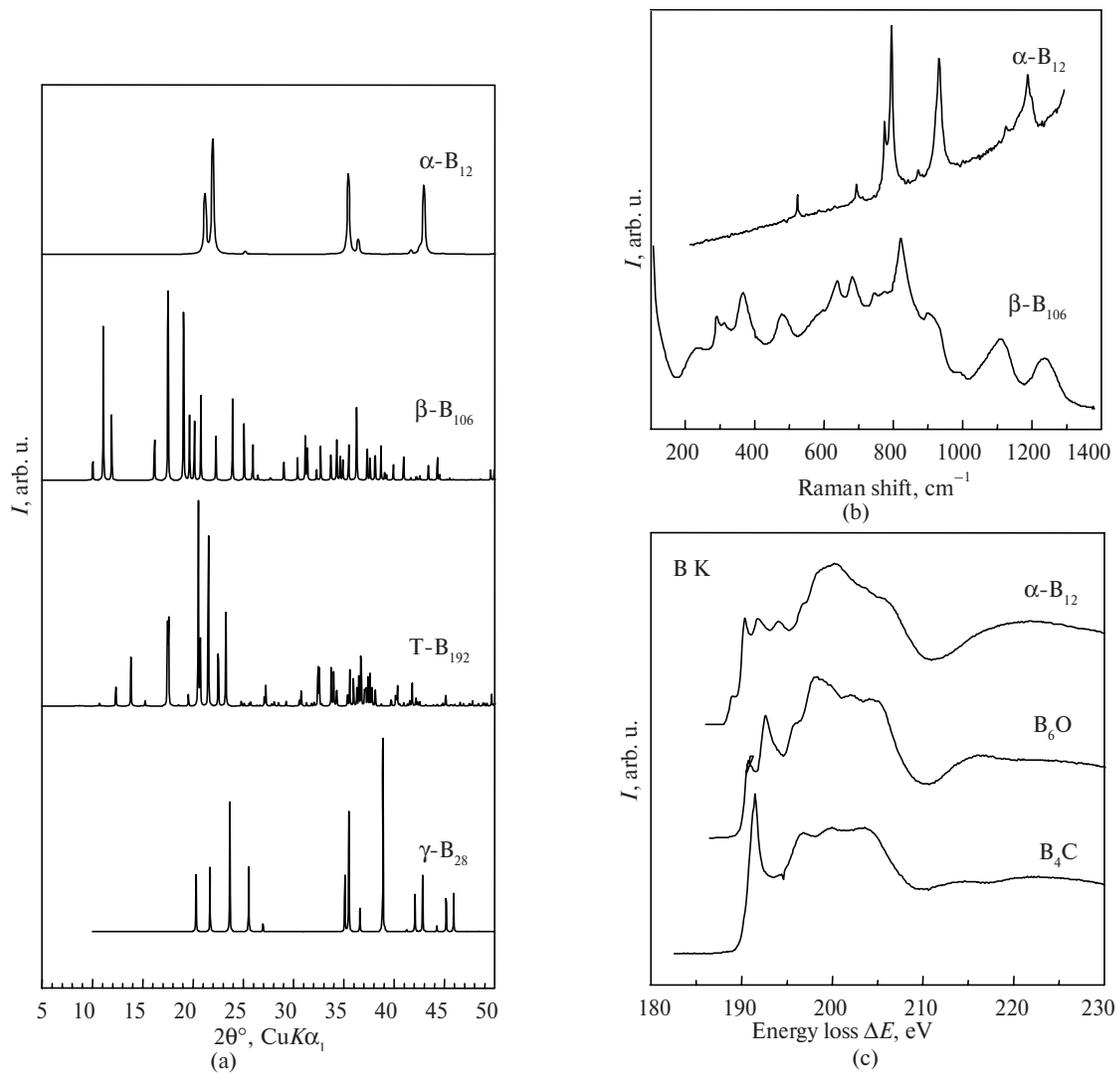

**Fig. 3.** XRD patterns of four known boron modifications (a), Raman spectra of α-$B_{12}$ and β-$B_{106}$ modifications (b), and EELS spectra of boron atoms in compounds with the α-$B_{12}$ structure (c).

Tetragonal boron T-$B_{192}$ (see Fig. 2d) [73] crystallizes in the $P4_3m$ space group (tetragonal syngony, $a =$ 10.14 Å, $c = 14.17$ Å) and belongs to the α-$AlB_{12}$ structural type. Its XRD pattern is given in Fig. 3a. Up to now properties of this phase have not been comprehensively studied.

There is also in the literature a description of hypothetical I-tetragonal boron phase T-$B_{52}$ of space group $P42/nmm$ or $P$-$42m$ (tetragonal syngony). This phase has never been obtained in the pure state and can be stabilized only by nitrogen or carbon impurities (3.85 at %) in the case of CVD synthesis at atmospheric pressure [81]. Recent results allow a suggestion that at 5 GPa this phase stabilizes at nitrogen impurities in the amount an order of magnitude lower than in the case of stabilization at near atmospheric pressure [82].

**Table 2.** Properties of some boron phases

| Properties | α-$B_{12}$ | β-$B_{106}$ | γ-$B_{28}$ |
|---|---|---|---|
| Hardness $H_V$, GPa | 42 | 49; 40; 34 | 50 [34] |
| Hardness $H_K$, GPa |  | 45 |  |
| Fracture toughness $K_{Ic}$, MPa·m$^{1/2}$ |  | 3 |  |
| Bulk modulus $B_0$, GPa | 207 [83]; 229 [84] | 210 [85] | 237 [86] |
| Pressure derivative of $B_0$, $B_0'$ | 4.2 [83]; 4 [84] | 2.23 [85] | 2.7 [86] |
| Density ρ, g·cm$^{-3}$ | 2.447 | 2.280 | 2.544 |

The β-$B_{106}$ phase is thermodynamically stable under the normal conditions [46], but even small amount of other light elements as impurities stabilizes other structural types of boron (mainly α-$B_{12}$ and T-$B_{52}$). High pressures also contribute to the stabilization of structures of types simpler than β-$B_{106}$.

## 4. TWO-ELEMENT PHASES OF THE B–N SYSTEM

At present many polymorphous modifications of boron nitride (BN) are systematically studied. In the general case their properties are similar to those of cubic (cBN) or hexagonal (hBN) boron nitride.

Cubic boron nitride, cBN, is the second to diamond superhard material in terms of industry. It crystallizes in the *F*-43*m* space group (cubic syngony, *a* = 3.6150 Å) and its structure is shown in Fig. 4a. The basic polytype of cBN is wurtzitic boron nitride (wBN), which is close to lonsdaleite (polytype of diamond) in structure and whose mechanical properties resemble those of cBN. The XRD pattern of cBN exhibits the 111, 220, 311, and 400 Bragg reflections, like that of diamond, and in addition, the 200 and 222 lines typical of structures of the sphalerite type (Fig. 5a). For cBN the threefold degenerate mode of irreducible representation $\Gamma'_{15}$ (or $T_2$) is active in the Raman spectrum. This optical mode splits into longitudinal and lateral modes due to the ionic nature of the B–N bond (1055 and 1304 cm$^{-1}$). Doping cBN with other elements causes a disorder in the structure and as a result brings about the appearance of other phonon modes in the spectrum [87]. EELS spectra of cBN exhibit two signals corresponding to nitrogen and boron in a tetrahedral coordination [47, 48] (see Figs. 5b, 5c). The cBN basic properties are listed in Table 3.

Hexagonal boron nitride was first synthesized in the mid-nineteenth century [88]. However, it became a commercial product only a century later. hBN (2H polytype) crystallizes in the *P*-6*m*2 space group (hexagonal syngony, *a* = 2.5040 Å, *c* = 6.6612 Å) (see Fig. 4b). The basic polytype of hBN is rhombohedral boron nitride rBN (3R). XRD pattern for hBN differs slightly from graphite because of different mutual ordering of layers [89] (see Fig. 5a). Two modes of irreducible representation of the center of the Brillouin zone $E_{2g}$ are active in the Raman spectrum (55 and 1364 cm$^{-1}$) [90]. EELS spectra typical for boron and nitrogen atoms in the $sp^2$ hybridization allow one to identify a graphite-like structure [47, 48] (see Figs. 5b, 5c). Basic properties of hBN are given in Table 3.

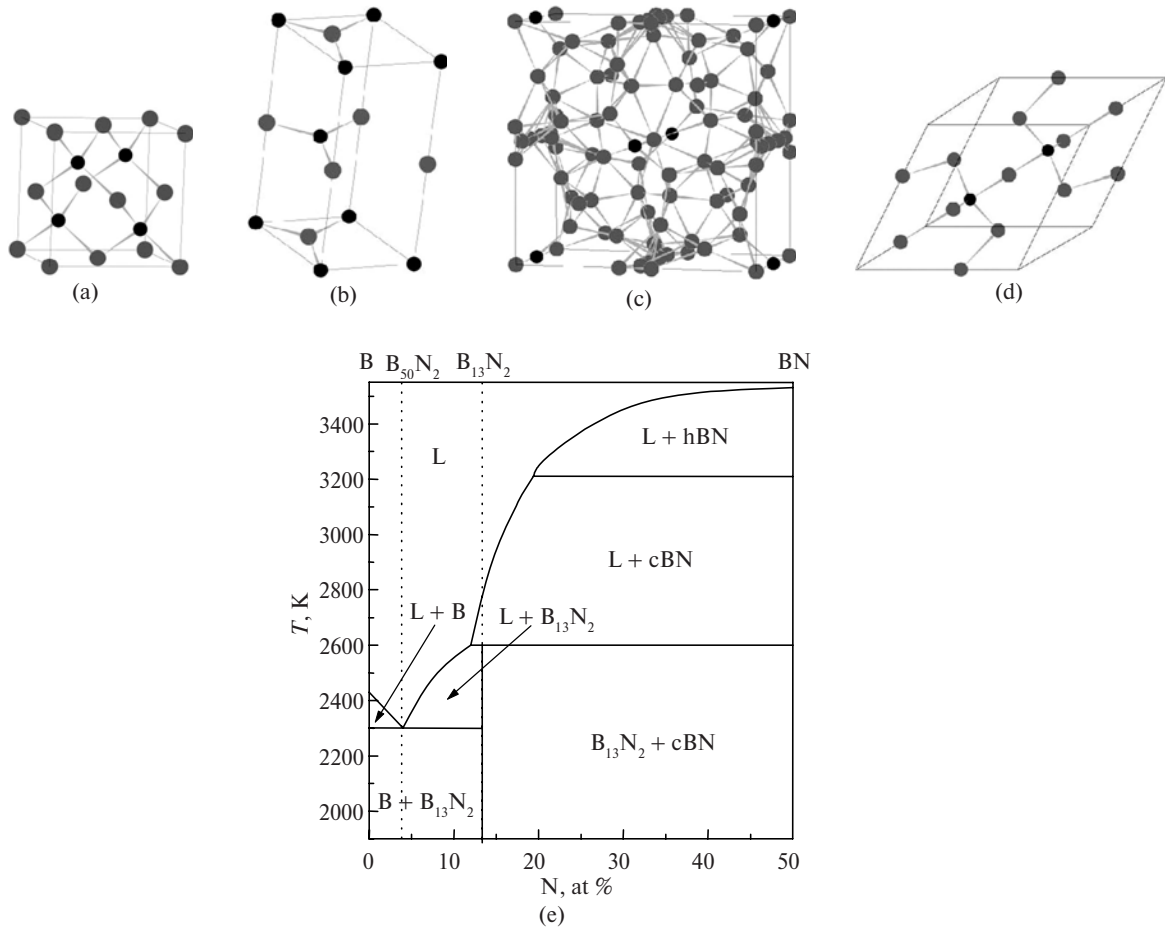

**Fig. 4.** Structures of phases of the B–N system: cBN (a), hBN (b), $B_{50}N_2$ (c), and $B_{13}N_2$ (d). Black balls indicate boron atoms, grey ones indicate nitrogen atoms. Equilibrium phase diagram of the B–BN system at 5 GPa (e).

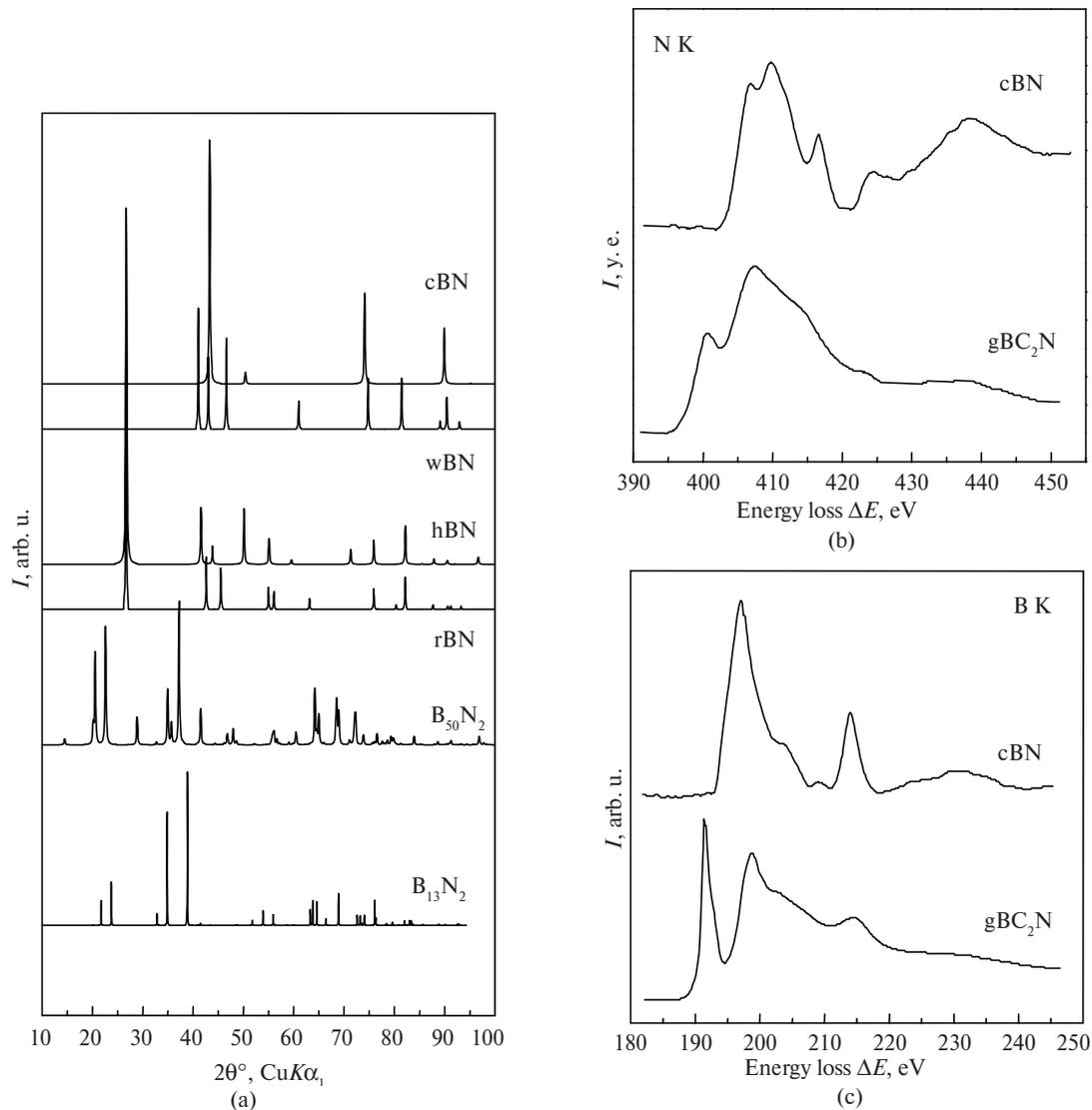

**Fig. 5.** XRD patterns of basic known phases of the B–N system (a), EELS spectra of nitrogen (b) and boron (c) atoms in $sp^2$ and $sp^3$ hybridizations (diamond- and graphite-like compounds).

For a long time $B_{50}N_2$ was the only subnitride, whose composition and structure were reliably established. It crystallizes in the *P42/nnm* or *P-42m* space group (tetragonal syngony, $a = 8.646$ Å, $c = 5.127$ Å) with a structure of the T-$B_{52}$ type (see Fig. 4c), which stabilizes by electron donors, i.e. atoms of nitrogen. This phase was first prepared by Laubengayer et al.[91] by CVD method and its structure was studied in later works [92, 93]. Its physicochemical properties have not, however, been studied comprehensively up to now.

The existence of the rhombohedral boron subnitride with a structure of the α-$B_{12}$ type was the subject of discussion for a long time. For the first time a similar phase of proposed composition $B_6N$ was described by Condon et al. [94] in the paper on kinetics of the boron and nitrogen interaction. Later the CVD synthesis of a phase of the same structural type of proposed composition $B_4N$ was reported in [95].

However, no evidence of the composition or structure was given in the papers. Hubert et al. [96] were the first to describe a solid-phase synthesis of subnitride of composition $B_6N$ produced by a reaction between amorphous boron and hBN at a pressure of 7.5 GPa and a temperature of 1700°C. However, as was shown in [97] the XRD pattern given in [96] could not belong to the proposed structure of the α-$B_{12}$ type. Thus, at the moment the nature of the described compounds is not ascertained yet, though in all probability they are subnitrides (either pure or with oxygen impurities).

Rhombohedral boron subnitride $B_{13}N_2$ was first synthesized from the B–BN melt by quenching at a pressure of 5 GPa and temperature 2400–2500 K and characterized by Solozhenko and Kurakevych [36, 98]. Subnitride crystallizes in the $R$-$3m$ space group (trigonal syngony, $a = 5.4455$ Å, $c = 12.2649$ Å). A rhombohedral unit cell is presented in Fig. 4d. An X-ray powder diffraction pattern is shown in Fig. 5a, basic properties are listed in Table 3. A Raman spectrum (Fig. 6a) is similar to those of boron suboxide and carbide, which is in full agreement with the data on the structures. The phase diagram of the B–BN system including $B_{13}N_2$ (see Fig. 4e) has been recently studied in situ at 5 GPa [99].

**Table 3.** Properties of some two-element phases of the B–N system

| Properties | cBN | $B_{13}N_2$ | hBN |
|---|---|---|---|
| Hardness $H_V$, GPa | 62 [26] | 40 [43] | |
| Hardness $H_K$, GPa | 44 [26] | | |
| Fracture toughness $K_{Ic}$, MPa·m$^{1/2}$ | 3.0 [26]; 6.8 [26] | | |
| Bulk modulus $B_0$, GPa | 377 [100]; 395 [101] | 200 [98] | 36.7 [102] |
| Pressure derivative of $B_0$, $B_0'$ | 4.1 [100]; 3.62 [101] | 4 [98] | 5.2 [102] |
| Young modulus, $E$, GPa | 909 | | |
| Density $\rho$, g·cm$^{-3}$ | 3.489 | 2.666 | 2.279 |

## 5. TWO-ELEMENT PHASES OF THE B–C SYSTEM

The majority of phases of the B–C system exhibit noticeably higher resistance to oxidation and interaction with iron group metals than corresponding carbon-based materials, which causes a growing interest in phases of this system.

Boron carbide $B_4C$ (in fact it is solid solution $B_{4+x}C_{1-x}$ having a wide concentration region of stability [35, 103], Fig. 7e) is a very hard substance, which may be produced at atmospheric pressure [104]. $B_4C$ crystallizes in the $R$-$3m$ space group (trigonal syngony, $a = 5.633$ Å, $c = 12.164$ Å) (see Fig. 7a). The Raman spectrum (Fig. 6a) of boron carbide is rather complicated as is the case with other phases having boron structure (it is similar to Raman spectrum of $\alpha$-$B_{12}$, just to this structural type it belongs). Active modes of $B_4C$ correspond to irreducible representation $6E_g + 5A_{1g}$ of the center of Brillouin zone. Typical EELS spectra for carbon and boron atoms in $B_4C$ are shown in Fig. 6b [76]. The phase basic properties are given in Table 4.

$B_{50}C_2$ is another known boron carbide that crystallizes in the $P42/nnm$ or $P$-$42m$ space group (tetragonal syngony, $a = 8.753$ Å, $c = 5.093$ Å) and has the T-$B_{52}$ type structure (see Fig. 7b), which stabilizes by electron donors, i.e. atoms of carbon. By now most of properties of this carbide have not been described in the literature.

The equilibrium solubility of boron in graphite has been reported in [105] and its maximum value is 2.3 at % at 2620 K (see Fig. 6c). A great number of metastable graphite-like phases of the B–C system with various stoichiometries (up to 50 at % B) were CVD synthesized [22, 23]. Turbostratic graphite-like phases t$BC_x$ (see Fig. 7c) have attracted interest because of their potential ability to be precursors for the synthesis of novel diamond-like phases of the B–C system [26, 27], which have semiconducting properties and are stable to oxidation [106, 108]. The XRD pattern exhibits virtually symmetrical lines 001 and 002 as well as two-dimensional asymmetrical reflections 10 and 11, which are typical of turbostratic graphite-like lattices [54]. The distribution of boron atoms in graphite-like layers is not known up to now, though the attempts to study the distribution using NMR have shown the existence of the B–B bonds at sufficiently high boron concentrations [23]. Under normal conditions the t$BC_x$ lattice parameters may vary in the range from 2.44 to 2.48 (the $a$ parameter) and from 3.40 to 3.44 (the $c$ parameter) and depend complicatedly on the boron concentration and CVD temperature [23, 109]. The EELS spectra exhibit typical signals of B and C atoms in the sp$^2$ hybridization [110], while the Raman spectra are identical to those of turbostratic graphite [111]. At high pressures and temperatures the t$BC_x$ phase segregation into boron-doped diamond and boron carbides takes place [29, 108, 112, 113], while at normal pressure disordered borated graphite forms instead of diamond. Thus, in the decomposition of graphite-like $BC_3$ at a pressure of 20 GPa and a temperature of 2200 K in a multi-anvil two-stage apparatus a conductive composite with an extreme hardness ($H_V$ = 92 GPa) [108] forms, while in the case of higher-boron precursors the hardness of resultant composites is several times lower [113].

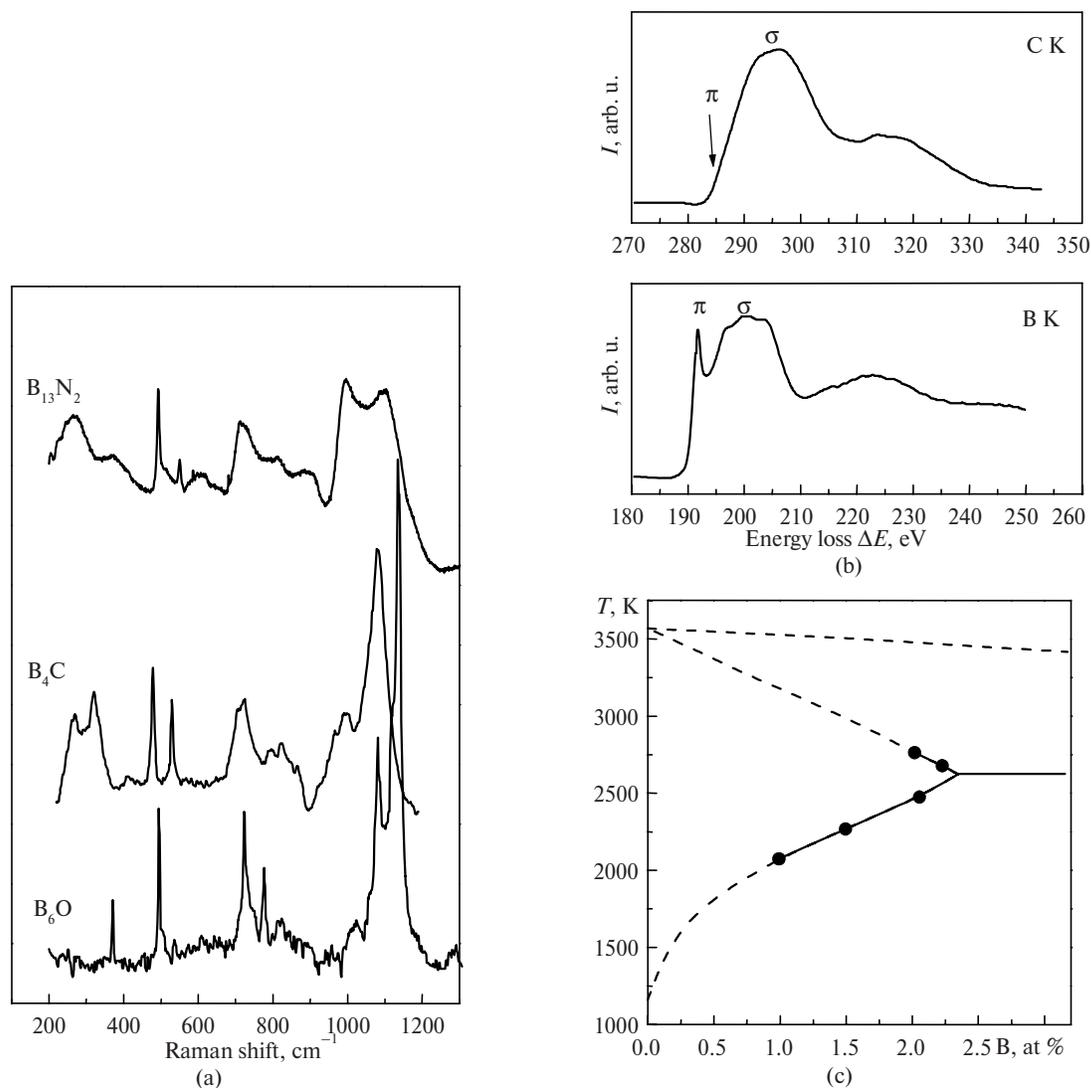

**Fig. 6.** Raman spectra of phases having the structures of the α-$B_{12}$, $B_{13}N_2$, $B_4C$, and $B_6O$ (a), EELS spectra of carbon and boron atoms in boron carbide (b), and the diagram of boron solubility in graphite (c).

Electron-deficient boron impurities in the structure strongly affect properties of respective carbon phases. Thus, boron-doped diamond is a semiconductor (and with sufficiently high boron concentrations the conductivity may become metallic), while diamond in itself is an insulator [13]. It has been supposed before that boron-doped diamond is a superconductor with a transition temperature of $T_c$ ~ 4 K [107], while for high-boron diamond-like $BC_x$ transition temperatures of about 55 K were predicted [114, 115]. At present it has been found that superconductivity of boron-doped diamond is caused not by boron in the lattice but by boron-containing grain boundaries of unknown phase composition [116].

Quite recently Solozhenko et al. have established that pseudocubic diamond-like phase of composition $BC_5$ ($Fd3m$ space group, cubic syngony, $a$ = 3.635 Å) (see Fig. 7d) [29] correspond to concentration limit of the existence of metastable boron-doped diamond. The structure and composition of this phase were defined using XRD, Raman, and EELS methods. The $cBC_5$ phase synthesized at pressures of 20–24 GPa and a temperature of 2200 K as a macroscopic single phase sample made possible a detail examination of its properties. It has been shown [29] that the $cBC_5$ phase is superhard and conducting, and is stable at temperatures, which exceed the temperature of diamond stability by 500 K. The basic properties of $cBC_5$ are given in Table 4. A number in situ studies using X-ray diffraction of synchrotron radiation and Raman spectroscopy under pres-sure have allowed one to define the mechanism of the transformation of turbostratic phases of the B–C–N system into diamond-like [29, 111, 117, 118].

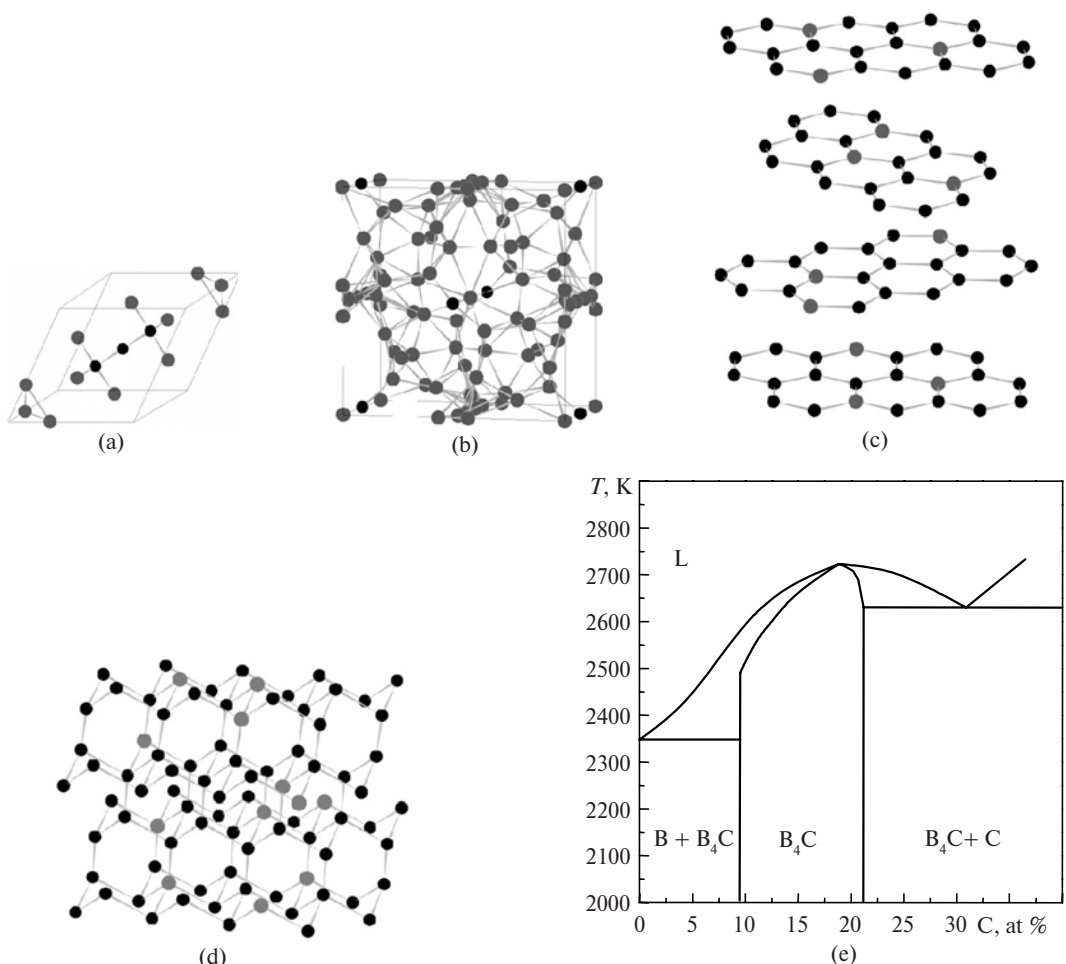

**Fig. 7.** Structures of phases of the B–C system: $B_4C$ (a), $B_{50}C_2$ (b), T-$BC_5$ (c), and c$BC_5$ (d) (black balls indicate carbon atoms, grey balls indicate boron atoms); phase diagram of the B–C system (e).

**Table 4.** Properties of some two-element phases of the B–C system

| Properties | $B_4C$ | c$BC_5$ | t$BC_x$ |
|---|---|---|---|
| Hardness $H_V$, GPa | 45; 38 [119] | 71 [29] | |
| Fracture toughness $K_{Ic}$, MPa·m$^{1/2}$ | 3–4 [120] | 9.5 [29] | |
| Bulk modulus $B_0$, GPa | 245 [119]; 199 [121] | 335 [29] | 20–30 [122] |
| Pressure derivative of $B_0$, $B_0'$ | 1 [121] | 4.5 [29] | 8–12 [122] |
| Density $\rho$, g·cm$^{-3}$ | 2.487 | 3.267 | |

## 6. TWO-ELEMENT PHASES OF THE B–O SYSTEM

$B_2O_3$ is an ordinary boron oxide, which is widely used as the initial compound for a synthesis of other phases containing boron and oxygen. Usually this oxide is met in the form of amorphous glass and crystalline α-$B_2O_3$ and β-$B_2O_3$. The α-$B_2O_3$ low-pressure phase crystallizes in the $P31$ space group (hexagonal syngony, $a = 4.3358$ Å, $c = 8.3397$ Å) [123] (Fig. 8a). At pressures above 2 GPa and high temperatures there forms the α-$B_2O_3$ high-pressure phase (orthorhombic syngony, the $Ccm2$ space group, $a = 4.613$ Å, $b = 7.803$ Å, $c = 4.129$ Å) [124] (see Fig. 8b). XRD patterns of both phases are shown in Fig. 9a. A phase diagram of $B_2O_3$ was first studied by Mackenzie and Claussen in quenching experiments [125] and later on it was revised by Brazhkin et al. on the strength of in situ experiments [126]. Recently it has been found that the hardness of the β-$B_2O_3$ high-pressure phase is similar to that of the WC–10% Co hard alloy [127]. The basic properties of β-$B_2O_3$ are given in Table 5.

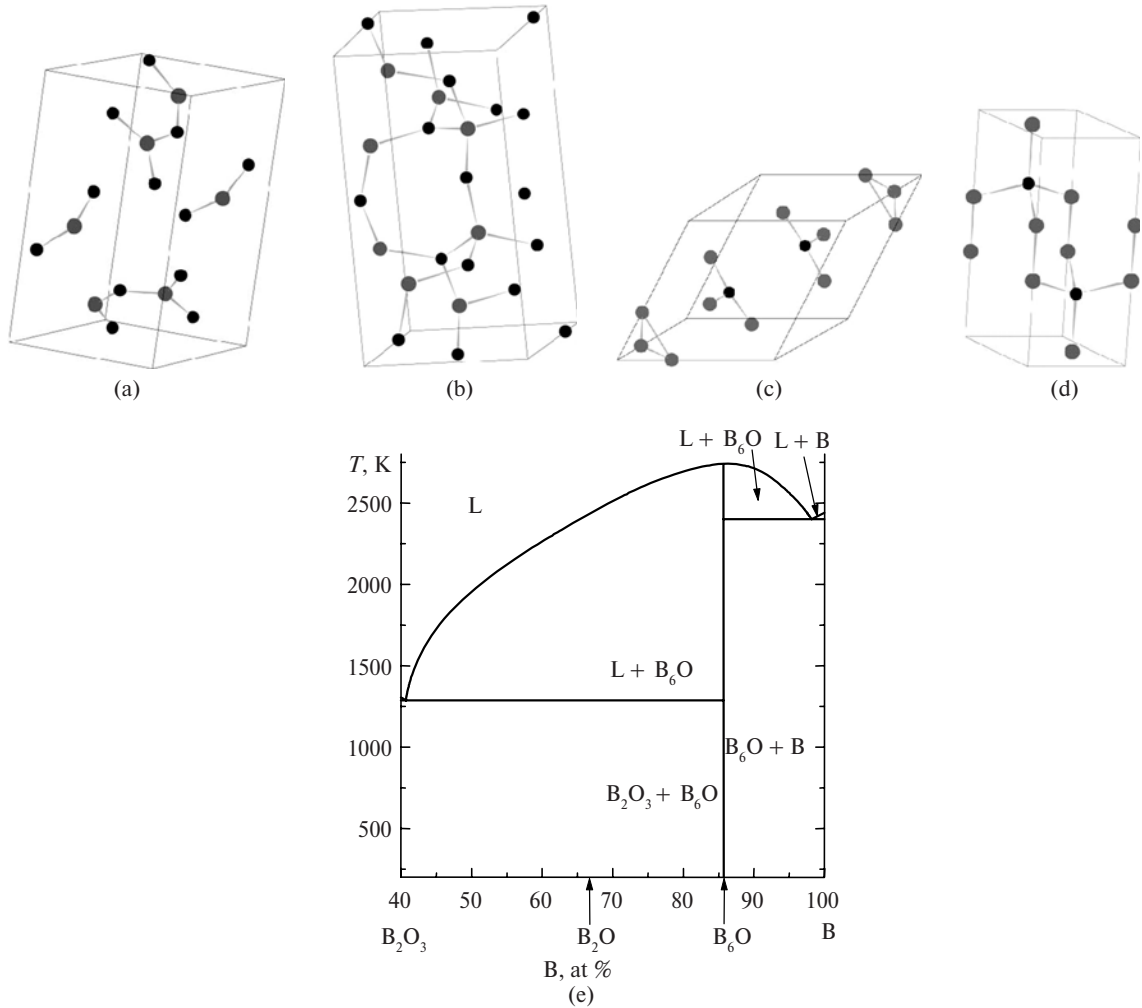

**Fig. 8.** Structures of phases of the B–O system: α-$B_2O_3$ (a), β-$B_2O_3$ (b), $B_6O$ (c), and hypothetical d$B_2O$ (d) (black balls indicate oxygen atoms, grey balls indicate boron atoms); phase diagram of the B–$B_2O_3$ system at 5 GPa (e).

Boron suboxide $B_6O$ has the structure of the α-$B_{12}$ type and crystallizes in the *R*-3*m* space group (trigonal syngony, $a = 5.3902$ Å, $c = 12.3125$ Å) [128, 129] (see Fig. 8c). XRD pattern is demonstrated in Fig. 9a. The Raman spectrum of this boron suboxide is rather complicated and cannot be excited by conventional green and red lasers due to strong fluorescence [103, 130]. Blue, IR, or UV laser excitation beams should be used to obtain the Raman spectrum. The active modes in the Raman spectrum correspond to irreducible representation $6E_g + 5A_{1g}$. Typical EELS spectra for $B_6O$ [76] are given in Fig. 9c. The suboxide basic properties are listed in Table 5. The procedure of the $B_6O$ synthesis was developed by Hubert et al. [129] and involves the interaction of amorphous boron with $B_2O_3$ at high pressures and temperatures. It has been found that a pressure of 4–5.5 GPa and temperature of 2000–2100 K are ideal conditions for the synthesis of a practically stoichiometric $B_6O$ of an icosahedral habit, though in the later papers it has been indicated that the synthesis of the practically stoichiometric phase is possible even at pressures above 1 GPa if crystalline β-$B_{106}$ is used instead of amorphous boron [130, 131]. At ambient pressure $B_6O$ decomposes above 2030 K to form boron and, possibly, $O_2$ [37]. The phase diagram of the B–$B_2O_3$ system including $B_6O$ at 5 GPa has been determined based on in situ experiments in the recent study of Solozhenko et al. [132] (see Fig. 8e).

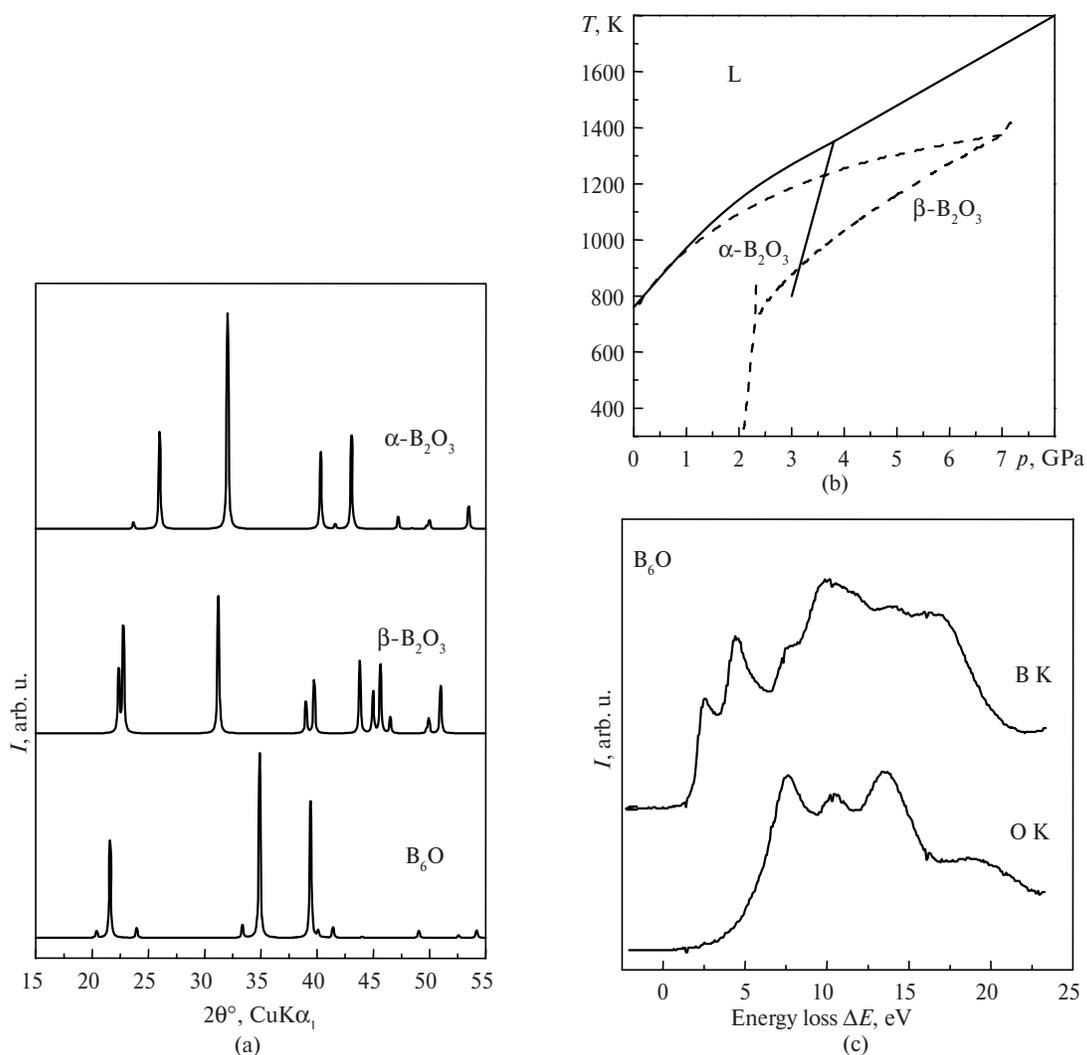

**Fig. 9.** XRD patterns from phases of the B–O system (a), phase diagram of $B_2O_3$ (b) and EELS spectra of oxygen and boron atoms in boron suboxide $B_6O$ (c).

The concept of symmetrical and asymmetrical isoelectronic analogs of carbon was put forward by Hall in 1965 [133]. In the same year he and Compton [134] reported the synthesis of graphite-like $B_2O$ ($gB_2O$) in the boron interaction with $B_2O_3$ at a pressure of 7.5 GPa and a temperature of 2100 K. Later Endo et al. [32] described the synthesis of diamond-like $B_2O$ ($dB_2O$), see Fig. 8d) by oxidation of boron phosphite BP using $CrO_3$ at 4 GPa and 1500 K and of graphite-like $gB_2O$ by oxidation of BP with $KClO_3$. However, none of these syntheses was reproduced later. It has been shown in [129, 135, 136] that $B_2O$ cannot be obtained by chemical interaction in the B–O system below 43 GPa and 2500 K. Moreover, it is indicated in [135] that $gB_2O$ described by Hall [133] is a mixture of $\beta$-$B_2O_3$ and undefined products of hydrolysis.

## 7. TWO-ELEMENT PHASES OF THE C–N SYSTEM

Graphite-like phases of the C–N system may be produced by the thermal decomposition of precursors of the C–N–Si–Cl–H and C–N–Si–F–H systems [21], reaction between melamine and cyanuric chloride [138], or polycondensation of melamine and supercritical hydrazine [139]. Despite the fact that the synthesized materials are of the $C_3N_4$ composition, the identification of their structures is ambiguous because of the heterogeneity of samples, which in most cases contain crystalline inclusions of an unknown composition in the $CN_x$ matrix. The synthesis of turbostratic carbon nitride (tCN) with a stoichiometry close to CN by the thermal decomposition of precursors of the C–N–H system at high pressure is described in [25]. The three-dimensional structure of this phase is close to that of turbostratic boron nitride tBN with a disordered sequence of layers of $C_3N_3$ hexagons [25, 140], while the structure of the layer can be described in the framework of the model proposed in [141].

Table 5. Properties of some two-element phases of the B–O system

| Properties | $B_6O$ | $\beta$-$B_2O_3$ |
|---|---|---|
| Hardness $H_V$, GPa | 38 [37]; 40–45 [137] | 16 [127] |
| Fracture toughness $K_{Ic}$, MPa·m$^{1/2}$ | 4.5 [137] | |
| Bulk modulus $B_0$, GPa | 181 [136] | 170 [136] |
| Pressure derivative of $B_0$, $B_0'$ | 6 [136] | 2.5 [136] |
| Density $\rho$, g·cm$^{-3}$ | 2.620 | 3.111 |

Theoretical studies [41, 45] were aimed at searching for phases, which would have high bulk modulus and high hardness. As a result the existence was suggested and properties were predicted of hypothetical $cC_3N_4$, $pcC_3N_4$, $\alpha$-$C_3N_4$, $\beta$-$C_3N_4$, and $hC_3N_4$ phases. For a long time graphite-like carbon nitride with the turbostratic structure has been considered as a promising precursor for obtaining $\beta$-$C_3N_4$ and/or other superhard forms of carbon nitride [41]. The synthesis of cubic $C_3N_4$ from turbostratic $tC_3N_4$ precursor in diamond anvils at pressures of 21–38 GPa and temperatures between 1600 and 3000 K has been recently reported [33]. However, the experimental data given in the paper do not allow one to distinguish the cubic lattice of $C_3N_4$ from the lattice of the NaCl high-pressure phase, which was used as the pressure medium.

## 8. CONCLUSIONS

The review describes the basic known superhard phases of the B–C–N–O system and basic precursors for the phase synthesis with a special emphasis on the structure and methods of characterization. The B–C–N–O quaternary system is still a promising object of investigation with the aim to search for novel superhard materials having unique properties for machining hard materials and the use in high-temperature electronics. Very high pressures and temperatures in combination with in situ observations by X-ray diffraction of synchrotron radiation are of particular importance for synthesis of these materials.

## 9. ACKNOWLEDGMENTS

The author thanks Dr. Sci. (Chem.) Solozhenko for the discussion and a number of critical comments and Prof. Novikov, Member of the National Academy of Sciences of Ukraine, for his support of this work.